\documentclass[aps,preprint]{revtex4-1} 
\usepackage{amsmath}    % need for subequations
\usepackage{graphicx}   % need for figures
\usepackage{verbatim}   % useful for program listings
\usepackage{color}      % use if color is used in text
\usepackage{subfigure}  % use for side-by-side figures
\usepackage{hyperref}   % use for hypertext links, including those to external documents and URLs
\usepackage{here}       % [H] for fixing position of a figure

\begin{document}

\preprint{KOBE-TH-15-05}
\title{Designing Anisotropic Inflation with Form Fields}
\author{Asuka Ito and Jiro Soda}
\affiliation{Department of Physics, Kobe University, Kobe 657-8501, Japan}

\date{\today}

\begin{abstract}
We study inflation with anisotropic hair induced by form fields. 
In four dimensions, the relevant form fields are gauge (one-form) fields and two-form fields.
Assuming the exponential form of potential and gauge kinetic functions, 
we find new exact power-law solutions endowed with anisotropic hair.
We also explore the phase space of anisotropic inflation and find fixed points corresponding to the exact power-law solutions.
 Moreover, we perform the stability analysis around the fixed points to reveal the structure of the phase space. 
It turns out that one of the fixed points becomes an attractor and others (if any) are saddle points. In particular, the one
corresponding to anisotropic inflation becomes an attractor when it exists.
We also argue that various anisotropic inflation models can be designed by choosing coupling constants.
\end{abstract}

\maketitle

%\tableofcontents

\section{Introduction}

 Scalar fields play a central role in inflationary cosmology.
 This is because a scalar field called an inflaton can mimic a cosmological constant and consequently drive a quasi-de Sitter inflation.  
 On the other hand, other fields such as gauge fields have been overlooked in the study of inflation.
 This is because the cosmic no-hair conjecture insists that other matter fields satisfying the dominant and strong energy conditions 
will  be rapidly diluted in the presence of the cosmological constant. 
Indeed, in the case of homogeneous universe, there exists the cosmic no-hair theorem~\cite{Wald:1983ky}.
Since the inflaton pretends to be a cosmological constant, it is natural to expect that  this theorem holds even in an inflationary universe. 
 Hence, it has been believed that gauge fields are irrelevant during inflation.
Nonetheless, there have been several attempts for finding a counter example to the cosmic no-hair conjecture by relaxing the energy conditions of matter
or by modifying gravity~\cite{Ford:1989me}. 
 It has been shown that almost all are unstable or need extreme fine tunings~\cite{Himmetoglu:2008zp}.
  However, recently, a clear counter example to the cosmic no-hair conjecture was found~\cite{Watanabe:2009ct,Kanno:2010nr}.
Indeed, there exists anisotropic inflation with a gauge field hair.  
The point was not to relax the energy conditions but to notice the slight difference between de Sitter and quasi-de Sitter spacetime.
Since the discovery of the anisotropic inflation, there have appeared many phenomenological 
applications~\cite{Gumrukcuoglu:2010yc} 
in conjunction with the statistical anisotropy~\cite{Soda:2012zm}.

Because of the phenomenological importance, it is worth clarifying to what extent this anisotropic inflation can be extended.
In other words, we need to design anisotropic inflation as broad as possible. 
 To this aim, we should note that the first counter example to the cosmic no-hair conjecture consists of a scalar field and a gauge field.
From the point of the cosmic democracy, however, it is natural to ask if other fields can survive during inflation.
Here, we do not consider models with higher order derivatives. Rather, we envisage an effective action dimensionally reduced from string theory.
As such a model, we consider the action
\begin{equation}
    S=\int d^{4}x \sqrt{-g}\left[ \frac{M_{p}^{2}}{2}R-\frac{1}{2}(\partial_{\mu}\phi)(\partial^{\mu}\phi)
                          - V(\phi) - \sum_{p=2}^4 \frac{1}{4}\bar{f}_{p}{}^{2}(\phi) F_{p}^2   \right]  \label{eq1} \ ,
\end{equation}
where $M_{p}$ represents the reduced plank mass, $g$ is the determinant of the metric, $R$ is the Ricci scalar,  $\phi$ can be regarded as
a zero-form field, $V(\phi)$ is a potential function, and $F_p$ are field strength of $(p-1)$-form fields coupled to an inflaton with gauge kinetic functions $\bar{f}_p (\phi)$. 
In four-dimensions, there are three possibilities, namely, gauge (one-form) fields, two-form fields, and three-form fields. However, three-form fields are
equivalent to the potential function. In fact, the field strength of three-form is four-form which must be proportional to the volume form.
Therefore, we do not need to consider this case.
 One may also think two-form fields are equivalent to scalar fields through the Hodge duality
$F_3 = *d\psi$ with a scalar field $\psi$. However, in the cosmological setup, we usually consider homogeneous field $F_{012}(t) = * \partial_3 \psi$
which depends only on a cosmic time. From the view point of a scalar field $\psi$, this is an inhomogeneous configuration
$\psi \propto x^3$,  although this is compatible with homogeneity of the universe.
 Actually, a more complicated version of this possibility is known as solid inflation~\cite{Bartolo:2013msa}. 
From the point of designing anisotropic inflation, however, it is convenient to regard two-form fields as fundamental fields~\cite{Ohashi:2013mka}.
In this general setup, we shall design anisotropic inflation models which exhibit interesting features.

In this paper, we study inflationary universe in the presence of form fields. 
In particular, we take exponential type potential and gauge kinetic functions.
In the case of the gauge field with this setup, we know there exists
exact power-law solutions~\cite{Kanno:2010nr,Yamamoto:2012tq,Hervik:2011xm}.  
Hence, we expect new exact solutions can be found in the presence of the two-form field.
Thus, the aim of this work is two-fold.
One is to find exact power-law solutions in the presence of both gauge and two-form fields.
The other is to reveal the structure of phase space of anisotropic inflation models.

Since the gauge field and the two-form field give rise to opposite anisotropy, inflation becomes 
isotropic for special coupling constants. While, in general, there arises anisotropy of the order of slow roll parameter.
 Remarkably, co-existence of the gauge field and the two-form field
gives rise to a complicated phase space structure. 
Thus, our models provide a variety of concrete framework incorporating
the statistical anisotropy testable by precise observations.

The organization of the paper is as follows.
In section II, we seek power-law solutions and obtain exact  power-law inflationary solutions with form fields which exhibit anisotropic expansion.
In section III, we perform dynamical system analysis and find fixed points in phase space. 
It turns out that there are four kind of fixed points corresponding to power-law solutions found in section II.
We clarify when those fixed points appear.
In section IV, we examine the linear stability of the fixed points. 
In section V, we classify the general solutions in terms of fixed points.
 It turns out  that only one fixed point becomes stable and the others are saddle points if any.
The attractor we obtain shows various anisotropy caused by the gauge field or the two-form field.
When both gauge field and two-form field survive, we find the convergence to the attractor is slow. 
We argue how anisotropic inflation can be designed by choosing coupling constants.
The final section is devoted to the conclusion.

\section{Exact power-law solutions}

In this section, we will seek exact power-law solutions driven by an inflaton $\phi$ 
in the presence of a gauge field $A_\mu$ and  a two-form field $B_{\mu\nu}$.
For this purpose, we will concentrate on the exponential type potential 
\begin{eqnarray}
V(\phi)= V_{0}e^{\lambda\frac{\phi}{M_{p}}} \label{exp-potential}
\end{eqnarray}
and gauge kinetic functions
\begin{eqnarray}
 \bar{f}_A (\phi) =  f_{A} \  e^{\rho_{A}\frac{\phi}{M_{p}}} \ ,  \qquad  
 \bar{f}_{B} (\phi) = f_{B} \  e^{\rho_{B}\frac{\phi}{M_{p}}}  \ ,
\end{eqnarray}
where $V_0$, $f_A$, $f_B$ are constants.
Here, coupling constants $\lambda, \rho_A , \rho_B$ characterize inflation models.
Now, we consider the following action
\begin{equation}
    S=\int d^{4}x \sqrt{-g}\left[ 
                     \frac{M_{p}^{2}}{2}R-\frac{1}{2}(\partial_{\mu}\phi)(\partial^{\mu}\phi)
                          - V(\phi) -\frac{1}{4}  \bar{f}_{A}{}^{2}(\phi) F_{\mu\nu}F^{\mu\nu}
                         -\frac{1}{12}\bar{f}_{B}{}^{2}(\phi) H_{\mu\nu\lambda} H^{\mu\nu\lambda}
                              \right] \ ,  \label{eq1}
\end{equation}
where the field strength tensors $F_{\mu\nu}$ and  $H_{\mu\nu\lambda}$ are defined by $F_{\mu\nu}=\partial_{\mu}A_{\nu}-\partial_{\nu}A_{\mu}$ and
 $H_{\mu\nu\lambda}=\partial_{\mu}B_{\nu\lambda}+\partial_{\nu}B_{\lambda\mu}+\partial_{\lambda}B_{\mu\nu}$ , respectively.
It is well known that there exists an isotropic power-law inflationary solution for the exponential type potential (\ref{exp-potential}).
In the presence of a gauge field, an anisotropic power-law inflationary solution has been found~\cite{Kanno:2010nr}. We shall find new exact solutions
in the presence of additional fields. 

Now, we are in a position to discuss cosmological homogeneous solutions.
First of all, without loss of generality, we can take the gauge field $A_{\mu}$ to be $A_{\mu}=(0,v_{A}(t),0,0)$.
Given this configuration, in principle, the two-form field can take an arbitrary polarization.
However, in the case of gauge fields, it has been shown that the fields tend to be orthogonal to make the expansion
as  isotropic as possible~\cite{Yamamoto:2012tq}.
 Thus, we assume  the two-form field $B_{\mu\nu}$ has an orthogonal configuration
 $\frac{1}{2}B_{\mu\nu}dx^{\mu}\wedge dx^{\nu}=v_{B}(t)dy\wedge dz$ to the gauge field.
Since these ansatzes keep the rotational symmetry in the $y$-$z$ plane, we can take the metric in the axially symmetric form
\begin{equation}
    ds^{2}=-dt^{2}+e^{2\alpha(t)}\left[\ e^{-4\sigma(t)}dx^{2}+e^{2\sigma(t)}(dy^{2}+dz^{2})\right] \ , 
\label{eq2}
\end{equation}
where $t$ is the cosmic time. Here, $\alpha$ describes the averaged expansion of  the universe 
and $\sigma$ represents anisotropy.  
From the action (\ref{eq1}), we can derive the equations of motion. 
It is easy to solve the equations of the gauge field and the two-form field as
\begin{eqnarray}
    \dot{v}_{A}=p_{A}f_{A}^{-2}e^{-2\rho_{A}\frac{\phi}{M_{p}}}e^{-\alpha-4\sigma} \ ,  \qquad
    \dot{v}_{B}=p_{B}f_{B}^{-2}e^{-2\rho_{B}\frac{\phi}{M_{p}}}e^{\alpha+4\sigma}  \ , 
   \label{integral}
\end{eqnarray}
where $p_{A},p_{B}$ are constants of integration.
Substituting these solutions into other equations, we can deduce the hamiltonian constraint
\begin{eqnarray}
    \dot{\alpha}^{2}=\dot{\sigma}^{2}+\frac{1}{3M_{p}^{2}}\left[\ \frac{1}{2}\dot{\phi}^{2}+V_{0}e^{\lambda\frac{\phi}{M_{p}}}+
                     \frac{1}{2}p_{A}^{2}f_{A}^{-2}e^{-2\rho_{A}\frac{\phi}{M_{p}}}e^{-4\alpha-4\sigma}
                     +\frac{1}{2}p_{B}^{2}f_{B}^{-2}e^{-2\rho_{B}\frac{\phi}{M_{p}}}e^{-2\alpha+4\sigma}\right] \ ,\quad
                   \label{eq3}
\end{eqnarray}
the rest of Einstein equations
\begin{eqnarray}
    \ddot{\alpha}&=&-3\dot{\alpha}^{2}+\frac{1}{M_{p}^{2}}V_{0}e^{\lambda \frac{\phi}{M_{p}}}+
                   \frac{1}{6M_{p}^{2}}p_{A}^{2}f_{A}^{-2}e^{-2\rho_{A}\frac{\phi}{M_{p}}}e^{-4\alpha-4\sigma}+
                   \frac{1}{3M_{p}^{2}}p_{B}^{2}f_{B}^{-2}e^{-2\rho_{B}\frac{\phi}{M_{p}}}e^{-2\alpha+4\sigma} \ ,
                   \label{eq4}\\
    \ddot{\sigma}&=&-3\dot{\alpha}\dot{\sigma}+
                  \frac{1}{3M_{p}^{2}}p_{A}^{2}f_{A}^{-2}e^{-2\rho_{A}\frac{\phi}{M_{p}}}e^{-4\alpha-4\sigma}
                  -\frac{1}{3M_{p}^{2}}p_{B}^{2}f_{B}^{-2}e^{-2\rho_{B}\frac{\phi}{M_{p}}}e^{-2\alpha+4\sigma} \ ,
                  \label{eq5}
\end{eqnarray}
and the equation of the scalar field
\begin{eqnarray}
    \ddot{\phi}&=&-3\dot{\alpha}\dot{\phi}-\frac{\lambda}{M_{p}} V_{0}e^{\lambda\frac{\phi}{M_{p}}}+
                \frac{\rho_{A}}{M_{p}}p_{A}^{2}f_{A}^{-2}e^{-2\rho_{A}\frac{\phi}{M_{p}}}e^{-4\alpha-4\sigma}
                +\frac{\rho_{B}}{M_{p}}p_{B}^{2}f_{B}^{-2}e^{-2\rho_{B}\frac{\phi}{M_{p}}}e^{-2\alpha+4\sigma} \ .
                \label{eq6}
\end{eqnarray}
To seek power-law solutions, we put the ansatz
\begin{eqnarray}
    \alpha=\zeta \log{M_{p}t}  \ ,\qquad
    \sigma=\eta \log{M_{p}t} \ , \qquad
    \frac{\phi}{M_{p}}&=&\xi \log{M_{p}t} +\phi_{0} \ ,
\label{eq11}
\end{eqnarray}
where $\zeta , \eta , \xi ,\phi_0$ are constants.
Substituting the above ansatz into the hamiltonian constraint (\ref{eq3}), we obtain
\begin{eqnarray}
    \frac{\zeta^{2}}{t^{2}}&=&\frac{\eta^{2}}{t^{2}}+\frac{1}{6}\frac{\xi^{2}}{t^{2}}+\frac{1}{3M_{p}^{2}}
             V_{0}e^{\lambda\phi_{0}}(M_{p}t)^{\lambda\xi}\nonumber \\ 
             &&+\frac{1}{6M_{p}^{2}}p_{A}^{2}f_{A}^{-2}e^{-2\rho_{A}\phi_{0}}(M_{p}t)^{- 2\rho_{A}\xi-4\zeta-4\eta}+
             \frac{1}{6M_{p}^{2}}p_{B}^{2}f_{B}^{-2}e^{-2\rho_{B}\phi_{0}}(M_{p}t)^{- 2\rho_{B}\xi-2\zeta+4\eta}
             \label{eq12} \ .
\end{eqnarray}
Taking a look at the power of the time $t$, we have the relations
\begin{eqnarray}
    \lambda \xi =-2&& \ , \label{relation-lambda} \\
    \rho_{A}\xi +2\zeta+2\eta&=&1 \label{A-relation} \ , \\
    \rho_{B}\xi +\zeta -2\eta&=&1 \label{B-relation} \ .
\end{eqnarray}
Here, we assumed form fields are non-trivial. In the case that the field vanishes, the corresponding relation is not necessary.
Let us define the following variables 
\begin{eqnarray}
    u=\frac{1}{M_{p}^{4}}V_{0}e^{\lambda\phi_{0}}    \ , \qquad
    w=\frac{1}{M_{p}^{4}}p_{A}^{2}f_{A}^{-2}e^{-2\rho_{A}\phi_{0}} \ ,\qquad
    z=\frac{1}{M_{p}^{4}}p_{B}^{2}f_{B}^{-2}e^{-2\rho_{B}\phi_{0}}  
                     \label{eq21}  \ .
\end{eqnarray}
Thus, from (\ref{eq12}), we have the relation
\begin{eqnarray}
    0=-\zeta^{2}+\eta^{2}+\frac{1}{6}\xi^{2}+\frac{1}{3}u+\frac{1}{6}w+\frac{1}{6}z\label{eq22} \ .
\end{eqnarray}
Similarly, from Eqs.(\ref{eq4}), (\ref{eq5}), and (\ref{eq6}), we can deduce the following relations
\begin{eqnarray}
    0&=&\zeta-3\zeta^{2}+u+\frac{1}{6}w+\frac{1}{3}z\label{eq23}  \ , \\
    0&=&\eta-3\zeta\eta+\frac{1}{3}w-\frac{1}{3}z\label{eq24}  \ ,  \\
    0&=&\xi-3\zeta\xi-u\lambda+w\rho_{A}+z\rho_{B}\label{eq25}  \ . 
\end{eqnarray}

\subsection{Isotropic power-law inflation}

First, we start with the well known power-law solution.
We consider the case that both of the gauge field and the 2-form field vanish,  $p_{A}=p_{B}=0$. 
Then, Eqs.(\ref{A-relation}) and (\ref{B-relation}) are not necessary.
Thus,  from (\ref{relation-lambda}),(\ref{eq22})$\sim$(\ref{eq25}), we easily obtain the isotropic power-law solution:
\begin{equation}
    \xi=-\frac{2}{\lambda},\ \ \ \zeta=\frac{2}{\lambda^{2}},\ \ \ \eta=0,\ \ \ u=\frac{2(6-\lambda^{2})}
           {\lambda^{4}}
    ,\ \ \ w=z=0  \ .
\label{eq26}
\end{equation}
Notice that we need the condition $\lambda^2 <6$ for the existence of the solution.
For sufficiently small $\lambda$, this describes inflationary universe.

\subsection{Anisotropic power-law inflation with  a gauge field}

Second, we consider  the case $p_A\neq 0 \ , p_{B}=0$.
In this case, it is known there exists an exact anisotropic power-law inflationary solution~\cite{Kanno:2010nr}.
The resultant solution is given by
\begin{eqnarray}
    \xi&=&-\frac{2}{\lambda}  \ , \label{eq46}\\
    \zeta&=&\frac{\lambda^{2}+8\rho_{A}\lambda+12\rho_{A}^{2}+8}{6\lambda(\lambda+2\rho_{A})}  \ ,\label{eq47}\\
    \eta&=&\frac{\lambda^{2}+2\rho_{A}\lambda-4}{3\lambda(\lambda+2\rho_{A})}    \ , \label{eq48}\\
    u&=&\frac{(2\rho_{A}^{2}+\rho_{A}\lambda+2)(-\lambda^{2}+4\rho_{A}\lambda+12\rho_{A}^{2}+8)}
              {2\lambda^{2}(\lambda+2\rho_{A})^{2}}  \ , \label{eq49}\\
    w&=&\frac{(\lambda^{2}+2\rho_{A}\lambda-4)(-\lambda^{2}+4\rho_{A}\lambda+12\rho_{A}^{2}+8)}
             {2\lambda^{2}(\lambda+2\rho_{A})^{2}}    \ , \label{eq50}\\
    z&=&0          \  .    \label{eq51}
\end{eqnarray}
From (\ref{eq50}),  we see the following condition is required for the existence of this power-law solution
\begin{equation}
    \rho_{A}>\frac{2}{\lambda}-\frac{1}{2}\lambda    \ .     \label{eq52}
\end{equation}
In inflationary universe $\lambda \ll 1$, this implies $\rho_{A}\gg 1$.

In order to see the consistency with  the cosmic no-hair theorem, we define the averaged slow roll parameter
\begin{equation}
    \epsilon \equiv -\frac{\dot{H}}{H^{2}} =
    \frac{6\lambda(\lambda+\rho_{A})}{\lambda^{2}+8\rho_{A}\lambda+12\rho_{A}^{2}+8}\ .\ 
    \label{slow-roll-1}
\end{equation}
The relation between  the averaged slow roll parameter and the anisotropy is 
given by
\begin{equation}
    \frac{\Sigma}{H}\equiv \frac{\dot{\sigma}}{\dot{\alpha}}=
    \frac{2(\lambda^{2}+2\rho_{A}\lambda-4)}{\lambda^{2}+8\rho_{A}\lambda+12\rho_{A}^{2}+8}=
    \frac{1}{3}I \epsilon \ ,
    \label{eq53}
\end{equation}
where we have defined
\begin{equation}
    I=\frac{\lambda^{2}+2\rho_{A}\lambda-4}{\lambda^{2}+\rho_{A}\lambda} \ .
\label{eq54}
\end{equation}
It should be stressed that the anisotropy vanishes in the limit $\epsilon \rightarrow 0$, which is consistent with the cosmic no-hair
theorem.

\subsection{Anisotropic power-law inflation with  a two-form field}

Next, we consider the case $p_{A}=0 \ , p_B \neq 0$.  
Then,  Eqs.(\ref{relation-lambda}),  (\ref{B-relation}) and (\ref{eq22})$\sim$(\ref{eq25}) can be written as
\begin{eqnarray}
    \lambda\xi&=&-2\label{eq27} \ ,\\
    \rho_{B}\xi+\zeta-2\eta&=&1\label{eq28} \ ,\\
    0&=&-\zeta^{2}+\eta^{2}+\frac{1}{6}\xi^{2}+\frac{1}{3}u+\frac{1}{6}z\label{eq29} \ ,\\
    0&=&\zeta-3\zeta^{2}+u+\frac{1}{3}z\label{eq30} \ , \\
    0&=&\eta-3\zeta\eta-\frac{1}{3}z\label{eq31}  \ , \\
    0&=&\xi-3\zeta\xi-u\lambda+z\rho_{B}\label{eq32} \ .
\end{eqnarray}
Eq.(\ref{eq27}) gives
\begin{equation}
    \xi=-\frac{2}{\lambda}\label{eq33} \ .
\end{equation}
From Eq.(\ref{eq31}), we have
\begin{equation}
    z=-3\eta(3\zeta-1)\label{eq34} \ .
\end{equation}
 Eqs. (\ref{eq30}) and  (\ref{eq34}) lead to
\begin{equation}
    u=(3\zeta-1)(\zeta+\eta)\label{eq35} \ .
\end{equation}
Using Eqs.(\ref{eq34}) and (\ref{eq35}) in Eq.(\ref{eq32}), we get
\begin{equation}
    (3\zeta-1)[\ \xi+\lambda(\zeta+\eta)+3\rho_{B}\eta\ ]=0 \label{eq36} \ .
\end{equation}
When $\zeta=\frac{1}{3}$, $u$ and $z$ become zero. This is not the solution which we desire. Hence, the solution should be
\begin{equation}
    \zeta=\frac{4+(2\rho_{B}+\lambda)(3\rho_{B}+\lambda)}{3\lambda(\lambda+\rho_{B})} \label{eq37} \ .
\end{equation}
%where we used Eqs.(\ref{eq33}) and (\ref{eq28}). 
Using this result in Eq.(\ref{eq28}), we obtain
\begin{equation}
    \eta=\frac{-\lambda^{2}-2\rho_{B}\lambda+2}{3\lambda(\lambda+\rho_{B})}\label{eq38} \ .
\end{equation}
Substituting the relations (\ref{eq37}) and (\ref{eq38}) into Eqs.(\ref{eq34}),(\ref{eq35}), we reach the solutions
\begin{eqnarray}
    u&=&\frac{(6\rho_{B}^{2}+4\rho_{B}\lambda+4)(2\rho_{B}^{2}+\rho_{B}\lambda+2)}{\lambda^{2}(\lambda+\rho_{B})^{2}}\label{eq39} \ , \\
    z&=&\frac{(\lambda^{2}+2\rho_{B}\lambda-2)(6\rho_{B}^{2}+4\rho_{B}\lambda+4)}{\lambda^{2}(\lambda+\rho_{B})^{2}} \ .
    \label{eq40}
\end{eqnarray}
It should be noted that Eq.(\ref{eq29}) is automatically satisfied.
Positivity of  Eq.(\ref{eq40}) requires the following condition
\begin{equation}
    \rho_{B}>\frac{1}{\lambda}-\frac{1}{2}\lambda \label{eq41} \ .
\end{equation}
If inflation occurs, $\lambda\ll1$,  the  above condition implies $\rho_{B}\gg1$.

For this solution, the averaged slow roll parameter is given by
\begin{equation}
    \epsilon =
    \frac{3\lambda(\lambda+\rho_{B})}{\lambda^{2}+5\rho_{B}\lambda+6\rho_{B}^{2}+4}\label{eq42} \ ,
\end{equation}
and the anisotropy is characterized by
\begin{equation}
    \frac{\Sigma}{H}=
    \frac{-\lambda^{2}-2\rho_{B}\lambda+2}{\lambda^{2}+5\rho_{B}\lambda+6\rho_{B}^{2}+4} 
           =-\frac{2}{3}I\epsilon   \ ,
    \label{eq43}
\end{equation}
where we defined
\begin{equation}
    I=\frac{\lambda^{2}+2\rho_{B}\lambda-2}{2\lambda(\lambda+\rho_{B})}   \ .\label{eq45}
\end{equation}
Again, this is consistent with the cosmic no-hair theorem.

\subsection{Anisotropic power-law inflation with a gauge field and a two-form field}

Finally, we consider the hybrid case that both of the gauge field and the 2-form field are non-trivial, $p_A\neq 0 \ ,  p_B \neq 0 $.
From Eqs.(\ref{relation-lambda})$\sim$(\ref{B-relation}), we have
\begin{eqnarray}
    \xi&=&-\frac{2}{\lambda}\label{eq55} \ , \\
    \zeta&=&\frac{2}{3}\frac{\lambda+\rho_{A}+\rho_{B}}{\lambda}\label{eq56} \ , \\
    \eta&=&\frac{1}{6}\frac{-\lambda+2(\rho_{A}-2\rho_{B})}{\lambda}\label{eq57}  \ .
\end{eqnarray}
Substituting  these results into Eqs.(\ref{eq23})$\sim$(\ref{eq25}), we obtain
\begin{eqnarray}
    u&=&\frac{1}{2}\frac{\lambda(2\rho_{A}+3\rho_{B})+2(2\rho_{A}^{2}+2\rho_{B}^{2}+\rho_{A}\rho_{B})+4}{\lambda^{2}}\label{eq58}\ ,\\
    w&=&\frac{\lambda^{2}-4+\lambda(2\rho_{A}-\rho_{B})+2\rho_{B}(\rho_{A}-2\rho_{B})}{\lambda^{2}}\label{eq59}\ ,\\
    z&=&\frac{1}{2}\frac{3\lambda^{2}-8+4\lambda(\rho_{A}+\rho_{B})-4\rho_{A}(\rho_{A}-2\rho_{B})}{\lambda^{2}}  \ .
\label{eq60}
\end{eqnarray}
The hamiltonian constraint (\ref{eq22}) is automatically satisfied. 
We should mention that the inflationary universe is isotropic in the special case, $\lambda=2(\rho_{A}-2\rho_{B})$. 
In this case,   the energy density of the gauge field 
is equal to that of the two-form field,  namely $w=z$.

\begin{figure}[!h]
\begin{center}
\includegraphics[width=9cm]{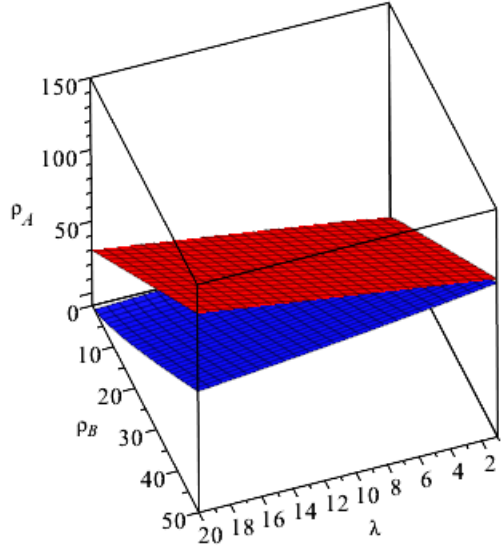}
\end{center}
\caption{We plotted  $\lambda^{2}-4+\lambda(2\rho_{A}-\rho_{B})+2\rho_{B}(\rho_{A}-2\rho_{B})=0$ as the blue curved surface above which is an allowed region.
 The red curved surface $3\lambda^{2}-8+4\lambda(\rho_{A}+\rho_{B})-4\rho_{A}(\rho_{A}-2\rho_{B})=0$ describes 
the upper bound of an allowed parameter region.
The hybrid solutions exist in the parameter region between the red and the blue curved surface.}
\label{Fig.1}
\end{figure}

Because of positivity of $w$ and $z $, Eqs.(\ref{eq59}) and (\ref{eq60}) yield the following inequalities
\begin{eqnarray}
    \lambda^{2}-4+\lambda(2\rho_{A}-\rho_{B})+2\rho_{B}(\rho_{A}-2\rho_{B})>0\label{eq61}  \ ,\\
    3\lambda^{2}-8+4\lambda(\rho_{A}+\rho_{B})-4\rho_{A}(\rho_{A}-2\rho_{B})>0\label{eq62} \  .
\end{eqnarray}
In FIG.\ref{Fig.1},  we plotted a parameter region for which the hybrid type solution exists. We find that as $\lambda$ becomes small, 
the region becomes narrow and approaches $\rho_{A} = 2\rho_B$ line. 

Now, the relation between the averaged slow roll parameter and the anisotropy is given by
\begin{equation} 
     \frac{\Sigma}{H}=\frac{1}{4}\frac{-\lambda+2(\rho_{A}-2\rho_{B})}{\lambda+\rho_{A}+\rho_{B}}=
             \frac{1}{6}I\epsilon\label{eq65}\ ,
\end{equation}
where
\begin{equation}
    \epsilon=\frac{3}{2}\frac{\lambda}{\lambda+\rho_{A}+\rho_{B}}\ ,\ \ \qquad
 I=\frac{-\lambda+2(\rho_{A}-2\rho_{B})}{\lambda}\label{eq66} \ .
\end{equation}
The consistency with the cosmic no-hair theorem is apparent.

\subsection{Anisotropic decelerating universe}

Note that we also have exact solutions which represent anisotropic decelerating universe.
The rate of acceleration is related to the slow roll parameter as
\begin{equation}
    \frac{\ddot{a}}{a}=H^{2}(1-\epsilon)\label{eq66.1}\ ,
\end{equation}
where $a$ is scale factor and $H= \dot{a}/{a}$ .
If $\epsilon>1$, the universe has decelerating expansion.
Moreover, the relation of the slow roll parameter and the parameter of the equation of state  $w$ is given by 
\begin{equation}
    w=\frac{2}{3}\epsilon-1\label{eq66.2} \ .
\end{equation}
For example, from Eq.(\ref{slow-roll-1}), we see $\epsilon$ can take the value close to 4 for $\lambda$ slightly above 4.
The corresponding equation of state is $w=5/3$. In the case of two-form dominated solutions, we have $\epsilon =3$ as an upper limit.
Thus, it describes anisotropic universe with $w=1$. For hybrid solutions, we have $w=0$ as an upper limit. 
This looks like an anisotropic matter  dominant universe.
So, we can describe the anisotropic decelerating universe with $-1/3<w<3$ by the power-law solutions.

\section{Fixed points in phase space}

In this section, we find fixed points in phase space.
It turns out that there are four kinds of fixed points corresponding to the power-law solutions we saw in the
previous section. 

Let us use the number of e-foldings $\alpha$ as a time coordinate.
It is convenient to define new variables
\begin{eqnarray}
    && X=\frac{\dot{\sigma}}{\dot{\alpha}}  \ ,\quad 
    Y=\frac{1}{M_{p}}\frac{\dot{\phi}}{\dot{\alpha}} \ ,\label{eq69}\\
   &&  Z_{A}=\frac{1}{M_{p}\dot{\alpha}}p_{A}f_{A}^{-1}e^{-\rho_{A}\frac{\phi}{M_{p}}}e^{-2\alpha-2\sigma} \ ,\quad
    Z_{B}=\frac{1}{M_{p}\dot{\alpha}}p_{B}f_{B}^{-1}e^{-\rho_{B}\frac{\phi}{M_{p}}}e^{-\alpha+2\sigma} \ .
\label{eq70}
\end{eqnarray}
With these variables, the equations of motion (\ref{eq4})$\sim$(\ref{eq6}) can be cast into an autonomous system.
Using Eqs.(\ref{eq4})$\sim$(\ref{eq6}), we can calculate the differential of (\ref{eq69})$\sim$(\ref{eq70}) with respect to $\alpha$
 as
\begin{eqnarray} 
    \frac{dX}{d\alpha}&=&X[\ 3(X^{2}-1)+\frac{1}{2}Y^{2}\ ]+\frac{1}{3}Z_{A}^{2}(X+1)+\frac{1}{6}Z_{B}^{2}(X-2) \ ,
    \label{eq71}\\
    \frac{dY}{d\alpha}&=&(Y+\lambda)[\ 3(X^{2}-1)+\frac{1}{2}Y^{2}\ ]+(\frac{1}{3}Y+\rho_{A}+
    \frac{\lambda}{2})Z_{A}^{2}
                         +(\frac{1}{6}Y+\rho_{B}+\frac{\lambda}{2})Z_{B}^{2}\label{eq72}  \ ,\\
    \frac{dZ_{A}}{d\alpha}&=&Z_{A}[\ 3(X^{2}-1)+\frac{1}{2}Y^{2}-\rho_{A}Y+1-2X+
                               \frac{1}{3}Z_{A}^{2}+\frac{1}{6}Z_{B}^{2}\ ]\label{eq73} \ ,\\
    \frac{dZ_{B}}{d\alpha}&=&Z_{B}[\ 3(X^{2}-1)+\frac{1}{2}Y^{2}-\rho_{B}Y+2+2X+
                               \frac{1}{6}Z_{B}^{2}+\frac{1}{3}Z_{A}^{2}\ ]\label{eq74}  \ .
\end{eqnarray}
The hamiltonian constraint (\ref{eq3}) is rewritten as
\begin{equation}
    -\frac{1}{M_{p}^{2}}\frac{V_{0}e^{\lambda \frac{\phi}{M_{p}}}}{\dot{\alpha}^{2}}=
        3(X^{2}-1)+\frac{1}{2}Y^{2}+\frac{1}{2}Z_{A}^{2}+\frac{1}{2}Z_{B}^{2}\label{eq75} \ .
\end{equation}
Because of the positivity of the potential of inflaton field, the above Eq.(\ref{eq75}) gives the condition
\begin{equation}
    0>3(X^{2}-1)+\frac{1}{2}Y^{2}+\frac{1}{2}Z_{A}^{2}+\frac{1}{2}Z_{B}^{2}\label{eq76} \ .
\end{equation}
We must seek fixed points which satisfy the above constraint.
To obtain fixed points, we solve Eqs. (\ref{eq71})$\sim$(\ref{eq74}) by setting the right-hand 
side to zero.
The solutions can be classified by $Z_A$ and $Z_B$.

\subsection{Isotropic fixed point: $Z_{A}=Z_{B}=0$}

First, we look for fixed points in the absence of form fields.
From Eq.(\ref{eq71}), we get
\begin{equation}
    X[\ 3(X^{2}-1)+\frac{1}{2}Y^{2}\ ]=0\label{eq77} \ .
\end{equation}
Here, $3(X^{2}-1)+\frac{1}{2}Y^{2}=0$ is against the constraint (\ref{eq76}), so we have
\begin{equation}
    X=0    \label{eq78}\ .
\end{equation}
Thus, from Eq.(\ref{eq72}), we obtain
\begin{equation}
    Y=-\lambda    \label{eq79}\ .
\end{equation}
This represents isotropic inflation and corresponds to the isotropic power-law solution which we saw in the 
previous section.

\subsection{ Gauge fixed point: $Z_{A}\neq 0,\ Z_{B}=0$}

Next, we take into account the gauge field~\cite{Kanno:2010nr}. 
Taking the same procedure as the isotropic case, we obtain
\begin{eqnarray}
    X&=&\frac{2(\lambda^{2}+2\rho_{A}\lambda-4)}{\lambda^{2}+8\rho_{A}\lambda+12\rho_{A}^{2}+8}\label{eq86} \ ,\\
    Y&=&\frac{-12(\lambda+2\rho_{A})}{\lambda^{2}+8\rho_{A}\lambda+12\rho_{A}^{2}+8}\label{eq87} \ ,\\
    Z_{A}^{2}&=&\frac{18(\lambda^{2}+2\rho_{A}\lambda-4)(-\lambda^{2}+4\rho_{A}\lambda+12\rho_{A}^{2}+8)}
                 {(\lambda^{2}+8\rho_{A}\lambda+12\rho_{A}^{2}+8)^{2}}\label{eq88} \ .
\end{eqnarray}
Note that from (\ref{eq88}), this fixed point requires the condition 
$\rho_{A}>\frac{2}{\lambda}-\frac{1}{2}\lambda$\ \ 
for the solution to exist. And this condition is the same as the condition (\ref{eq52}). Also this fixed point corresponds to the 
power-law solution with the gauge field.

\subsection{ Two-form  fixed point: $Z_{B}\neq 0,\ Z_{A}=0$}

Here, we consider the two-form field.
From  Eqs.(\ref{eq71}) and (\ref{eq72}), we have
\begin{equation}
    Y=-(\lambda +3\rho_{B})X-\lambda    \label{eq80} \ .
\end{equation}
Eq.(\ref{eq71}) gives 
\begin{equation}
    Z_{B}^{2}=\frac{-6X[\ 3(X^{2}-1)+\frac{1}{2}Y^{2}\ ]}{X-2}    \label{eq81} \ .
\end{equation}
Substituting solutions (\ref{eq80}),(\ref{eq81}) into Eq.(\ref{eq74}), we obtain
\begin{equation}
    (X+1)[\ (\lambda^{2}+5\rho_{B}\lambda+6\rho_{B}^{2}+4)X+\lambda^{2}+2\rho_{B}\lambda-2\ ]=0    \label{eq82} \ .
\end{equation}
Apparently, $X=-1$ is against the constraint (\ref{eq76}), so we take
\begin{equation}
    X=\frac{-\lambda^{2}-2\rho_{B}\lambda+2}{\lambda^{2}+5\rho_{B}\lambda+6\rho_{B}^{2}+4}    \label{eq83} \ .
\end{equation}
At this time, we have
\begin{eqnarray}
    Y&=&\frac{-6(\lambda+\rho_{B})}{\lambda^{2}+5\rho_{B}\lambda+6\rho_{B}^{2}+4}\label{eq84}  \ ,\\
    Z_{B}^{2}&=&\frac{9(\lambda^{2}+2\rho_{B}\lambda-2)(6\rho_{B}^{2}+4\rho_{B}\lambda+4)}  
               {\lambda^{2}+5\rho_{B}\lambda+6\rho_{B}^{2}+4}       \ .    \label{eq85}
\end{eqnarray}
Note that,  from the solution  (\ref{eq85}), we see 
$\rho_{B}$ must satisfy the condition $\rho_{B}>\frac{1}{\lambda}-\frac{1}{2}\lambda$  for 
the existence of this fixed point.  And this condition is the same as  the condition (\ref{eq41}).

\subsection{Hybrid fixed point:  $Z_{A}\neq 0,\ Z_{B}\neq 0$}

Finally, we consider the hybrid case.
From Eqs.(\ref{eq73}) and (\ref{eq74}), we have
\begin{equation}
    Y=\frac{4X+1}{\rho_{B}-\rho_{A}}    \ .
\label{eq89}
\end{equation}
Eq.(\ref{eq71}) gives
\begin{equation}
    Z_{A}^{2}=-\frac{1}{2}Z_{B}^{2}\frac{X-2}{X+1}-\frac{3X[\ 3(X^{2}-1)+\frac{1}{2}Y^{2}\ ]}{X+1}    \ .\label{eq90}
\end{equation}
Using Eqs.(\ref{eq90}) in (\ref{eq73}), we obtain
\begin{equation}
    Z_{B}^{2}=2X[\ 3(X^{2}-1)+\frac{1}{2}Y^{2}\ ]+2(X+1)[\ -3(X^{2}-1)-\frac{1}{2}Y^{2}+\rho_{A}Y-1+2X\ ] \ .
    \label{eq91}
\end{equation}
Substituting the solutions (\ref{eq89}),(\ref{eq90}) and (\ref{eq91}) into Eq.(\ref{eq72}), we get
\begin{equation}
    [\ (2\rho_{A}^{2}-3\rho_{A}\rho_{B}+\rho_{B}^{2}+4)X+2\rho_{A}^{2}-2\rho_{B}^{2}+1\ ]
    [\ (4\lambda+4\rho_{A}+4\rho_{B})X+\lambda-2\rho_{A}+4\rho_{B}\ ]=0    \label{eq92} \ .
\end{equation}
For $X=\frac{-2\rho_{A}^{2}+2\rho_{B}^{2}-1}{2\rho_{A}^{2}-3\rho_{A}\rho_{B}+\rho_{B}^{2}+4}$ , the 
constraint (\ref{eq76}) is not satisfied. So the fixed point we seek is as follows
\begin{eqnarray}
    X&=&\frac{1}{4}\frac{-\lambda+2(\rho_{A}-2\rho_{B})}{\lambda+\rho_{A}+\rho_{B}}\label{eq93} \ ,\\
    Y&=&-3\frac{1}{\lambda+\rho_{A}+\rho_{B}}\label{eq94}  \ , \\
    Z_{A}^{2}&=&\frac{9}{4}\frac{\lambda^{2}-4+\lambda(2\rho_{A}-\rho_{B})+2\rho_{B}(\rho_{A}-2\rho_{B})}
                 {(\lambda+\rho_{A}+\rho_{B})^{2}}\label{eq95}    \ , \\
    Z_{B}^{2}&=&\frac{9}{8}\frac{3\lambda^{2}-8+4\lambda(\rho_{A}+\rho_{B})-4\rho_{A}(\rho_{A}-2\rho_{B})}
                   {(\lambda+\rho_{A}+\rho_{B})^{2}}           \  .  \label{eq96}
\end{eqnarray}
The conditions stemming from the positivity of the $Z_{A}^{2}$ and $Z_{B}^{2}$ are the same as the conditions
(\ref{eq61}) and (\ref{eq62}). This fixed point corresponds to the hybrid power-law solution.

\section{Stability of fixed points}

In the previous section, we found the fixed points in phase space. According to the general theory of dynamical system,
the structure of phase space can be determined by the linear analysis around fixed points. 
Thus, in this section, we examine the linear stability of the fixed points which we found in the previous section.

Now, we linearize Eqs.(\ref{eq71})$\sim$(\ref{eq74}) around fixed points as
\begin{eqnarray}
    \frac{d(\delta X)}{d\alpha}&=&
              \delta X[\ 3(X^{2}-1)+\frac{1}{2}Y^{2}+6X^{2}+\frac{1}{3}Z_{A}^{2}+\frac{1}{6}Z_{B}^{2}\ ]\nonumber\\
              &&+\delta Y[\ XY\ ]+\delta Z_{A}[\ \frac{2}{3}(X+1)Z_{A}\ ]+
              \delta Z_{B}[\ \frac{1}{3}(X-2)Z_{B}\ ] \label{eq97}   \ , \\
    \frac{d(\delta Y)}{d\alpha}&=&
              \delta X[\ 6X(Y+\lambda)\ ]\nonumber \\
              &&+\delta Y[\ 3(X^{2}-1)+\frac{1}{2}Y^{2}+Y(Y+\lambda)+\frac{1}{3}Z_{A}^{2}+\frac{1}{6}Z_{B}^{2}\ ]
              \nonumber\\
              &&+\delta Z_{A}[\ 2Z_{A}(\frac{1}{3}Y+\rho_{A}+\frac{\lambda}{2})\ ]+
              \delta Z_{B}[\ 2Z_{B}(\frac{1}{6}Y+\rho_{B}+\frac{\lambda}{2})\ ]          \   , \label{eq98}\\
    \frac{d(\delta Z_{A})}{d\alpha}&=&\delta X[\ Z_{A}(6X-2)\ ]+\delta Y[\ Z_{A}(Y-\rho_{A})\ ]\nonumber \\
              &&+\delta Z_{A}[\ 3(X^{2}-1)+\frac{1}{2}Y^{2}-\rho_{A}Y+1-2X+Z_{A}^{2}+\frac{1}{6}Z_{B}^{2}\ ]
              \nonumber \\
              &&+\delta Z_{B}[\ \frac{1}{3}Z_{A}Z_{B}\ ]        \  ,  \label{eq99}\\
    \frac{d(\delta Z_{B})}{d\alpha}&=&\delta X[\ Z_{B}(6X+2)\ ]+\delta Y[Z_{B}(Y-\rho_{B})]+
              \delta Z_{A}[\ \frac{2}{3}Z_{A}Z_{B}\ ]\nonumber \\
              &&+\delta Z_{B}[\ 3(X^{2}-1)+\frac{1}{2}Y^{2}-\rho_{B}Y+2+2X+\frac{1}{2}Z_{B}^{2}+
              \frac{1}{3}Z_{A}^{2}\ ]   \ .
              \label{eq100}
\end{eqnarray}
To examine the stability of these equations, we set
\begin{eqnarray}
    \delta X=e^{\omega\alpha}\delta X'      \ , \quad
    \delta Y=e^{\omega\alpha}\delta Y'      \ , \quad
    \delta Z_{A}=e^{\omega\alpha}\delta Z_{A}'        \ , \quad
    \delta Z_{B}=e^{\omega\alpha}\delta Z_{A}'  \ .
  \label{eq109}
\end{eqnarray}
Then differential equations reduce to eigenvalue problem. 
Thus, what we need is to find the eigenvalues of the system. 
In the following, we discuss the stability of fixed points case by case.

\subsection{ Stability of isotropic fixed point}

Let us consider the stability of the isotropic fixed point.
Substituting the background solution  $Z_{A}=Z_{B}=0$, (\ref{eq78}) and (\ref{eq79}) into above linearized equations, we obtain
\begin{eqnarray}
    \frac{d(\delta X)}{d\alpha}&=&(\frac{1}{2}\lambda^{2}-3)\delta X \label{eq101}   \ ,\\
    \frac{d(\delta Y)}{d\alpha}&=&(\frac{1}{2}\lambda^{2}-3)\delta Y \label{eq102}  \ , \\
    \frac{d(\delta Z_{A})}{d\alpha}&=&(\frac{1}{2}\lambda^{2}+\rho_{A}\lambda-2)\delta Z_{A} \label{eq103}  \ ,\\
    \frac{d(\delta Z_{B})}{d\alpha}&=&(\frac{1}{2}\lambda^{2}+\rho_{B}\lambda-1)\delta Z_{B} \label{eq104}  \ .
\end{eqnarray}
The above equations are already diagonalized. Hence, it is easy to read off the eigenvalues.
Since $\lambda^2 <6 $, this fixed point is stable in the direction of $X$ and $Y$.
On the other hand, Eqs.(\ref{eq103}) and (\ref{eq104}) tell us that the stability requires
 $\rho_{A}<\frac{2}{\lambda}-\frac{1}{2}\lambda$\  and\  $
\rho_{B}<\frac{1}{\lambda}-\frac{1}{2}\lambda$. 
Taking look at the conditions (\ref{eq52}) and (\ref{eq41}), the isotropic fixed point is stable only in the absence of other fixed points. 

\subsection{ Stability of gauge fixed point}

Next, we examine the stability of the gauge fixed point~\cite{Kanno:2010nr}. The existence of this fixed point is guaranteed if
the condition  (\ref{eq52}) is satisfied.
The linearized equations with approximations $\lambda\ll1,\ \rho_{A}\gg 1$ become
\begin{eqnarray}
    \frac{d(\delta X)}{d\alpha}&=&-3\delta X  \ ,\\
    \frac{d(\delta Y)}{d\alpha}&=&-3\delta Y+\sqrt{6(\lambda^{2}+2\rho_{A}\lambda-4)} \delta Z_{A} \ , \\
    \frac{d(\delta Z_{A})}{d\alpha}&=&-\frac{1}{2}\sqrt{6(\lambda^{2}+2\rho_{A}\lambda-4)}\delta Z_{A} \ ,\\
    \frac{d(\delta Z_{B})}{d\alpha}&=&\frac{3(3\lambda^{2}+4\rho_{A}\lambda+4\rho_{B}\lambda-4\rho_{A}^{2}+
                                 8\rho_{A}\rho_{B}-8)}{\lambda^{2}+8\rho_{A}\lambda+12\rho_{A}^{2}+8}\delta Z_{B} \ .
\end{eqnarray}
Because of the condition  (\ref{eq52}), the square root in the above equations is real. 
Note that $X$ and $Z_B$ directions are already diagonalized. 
The eigenvalues are given by
\begin{equation}
    \omega=-3,\ -\frac{3}{2}\pm i\sqrt{3(\lambda^{2}+2\rho_{A}\lambda-4)-\frac{9}{4}},\ \ 
         \frac{3(3\lambda^{2}+4\rho_{A}\lambda+4\rho_{B}\lambda-4\rho_{A}^{2}+
                                 8\rho_{A}\rho_{B}-8)}{\lambda^{2}+8\rho_{A}\lambda+12\rho_{A}^{2}+8}   \ .
\label{eq111}
\end{equation}
\\
Apparently, the first three eigenvalues have negative real part. So this fixed point becomes stable if  $3\lambda^{2}+4\rho_{A}\lambda+4\rho_{B}\lambda-4\rho_{A}^{2}+8\rho_{A}\rho_{B}-8 < 0$. 
This condition is violated when the hybrid fixed point appears as one can see from Eq.(\ref{eq62}).

\subsection{ Stability of two-form  fixed point}

Here, we investigate the stability of the two-form fixed point.
Substituting solutions (\ref{eq83}),(\ref{eq84}) and (\ref{eq85}) into linearized equations, we obtain
\begin{eqnarray}
    \frac{d(\delta X)}{d\alpha}&=&-3\delta X\label{eq105} \ , \\
    \frac{d(\delta Y)}{d\alpha}&=&-3\delta Y+\sqrt{6(\lambda^{2}+2\rho_{B}\lambda-2)}\ \delta Z_{B}\label{eq106} \ , \\
    \frac{d(\delta Z_{B})}{d\alpha}&=&-\frac{1}{2}\sqrt{6(\lambda^{2}+2\rho_{B}\lambda-2)}\ \delta Y\label{eq107} \ ,
\end{eqnarray}
where we used approximations $\lambda\ll1,\ \rho_{B}\gg 1$ which is the case of inflation. 
The reality of square root is guaranteed by the condition (\ref{eq41}).
On the other hand, we exactly have
\begin{equation}
    \frac{d(\delta Z_{A})}{d\alpha}=\frac{3(\lambda^{2}+2\rho_{A}\lambda-\rho_{B}\lambda+
                                      2\rho_{A}\rho_{B}-4\rho_{B}^{2}-4)}{\lambda^{2}+5\rho_{B}\lambda
                                      +6\rho_{B}^{2}+4} \delta Z_{A}\label{eq108}  \ .
\end{equation}
The eigenvalues are given by
\begin{eqnarray}
    \omega=-3,-\frac{3}{2}\pm i\sqrt{3(\lambda^{2}+2\rho_{B}\lambda-2)-\frac{9}{4}},\ 
                                      \frac{3(\lambda^{2}+2\rho_{A}\lambda-\rho_{B}\lambda+
                                      2\rho_{A}\rho_{B}-4\rho_{B}^{2}-4)}{\lambda^{2}+5\rho_{B}\lambda+6\rho_{B}^{2}+4} \ . \quad \label{eq110}  
\end{eqnarray}
Again, first three eigenvalues have negative real part. So this fixed point becomes stable if $\lambda^{2}+2\rho_{A}\lambda-\rho_{B}\lambda+2\rho_{A}\rho_{B}-4\rho_{B}^{2}-4 < 0$. 
Again, this condition is violated when the hybrid fixed point appears as one can see from Eq.(\ref{eq61}).

\subsection{ Stability of hybrid fixed point}

Finally, we consider the stability of the hybrid fixed point.
The equations which we obtained by substituting solutions (\ref{eq93})$\sim$ (\ref{eq96}) into Eqs.(\ref{eq97}) $\sim$ (\ref{eq100}) 
are complicated. 
So this time, we give the results for several representative cases.
First, we fix two of the coupling constants as $\lambda=0.1,\ \rho_{B}=30$.
Then, the conditions (\ref{eq61}) and (\ref{eq62}) yield allowed values $59.9167<\rho_{A}<60.1168$.
Thus, we take parameter sets $(\lambda, \rho_{A}, \rho_{B})=(0.1, 59.9168, 30), (0.1, 60.0,30), (0.1, 60.1167, 30)$ 
for which the eigenvalues become
\begin{eqnarray}
    &&(-1.5\times 10^{-6}, \ -3.0, \ -1.5\pm 3.1i) \ ,\label{other-set0}\\
    &&(-2.3\times 10^{-3}, \ -3.0, \ -1.5\pm 3.8i) \ , \\
    &&(-2.0\times 10^{-6}, \ -3.0, \ -1.5\pm 4.7i) \  ,
\label{other-set}
\end{eqnarray}
respectively.
We have also checked other parameters and found the hybrid fixed point is stable.
Thus, we can infer the fixed point is always stable although we could not prove it analytically. 
Note that first eigenvalues in (\ref{other-set0})  $\sim$ (\ref{other-set}) are much less than 1.
Hence, the convergence of this direction is quite slow.

\section{Dynamical structure of phase space}

\begin{figure}[H]
\begin{center}
\includegraphics[width=8cm]{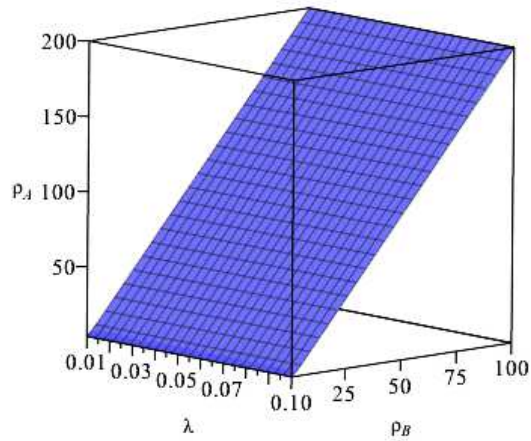}
\end{center}
\caption{In the parameter space $(\lambda , \rho_A , \rho_B )$, we depicted the blue  curved surface satisfying the relation 
 $\lambda^{2}+2\rho_{A}\lambda-\rho_{B}\lambda+2\rho_{A}\rho_{B}-4\rho_{B}^{2}-4 =0$.  Below this surface, the two-form field 
fixed point is stable. }
\label{Fig.2}
\end{figure}

\begin{figure}[H]
\begin{center}
\includegraphics[width=8cm]{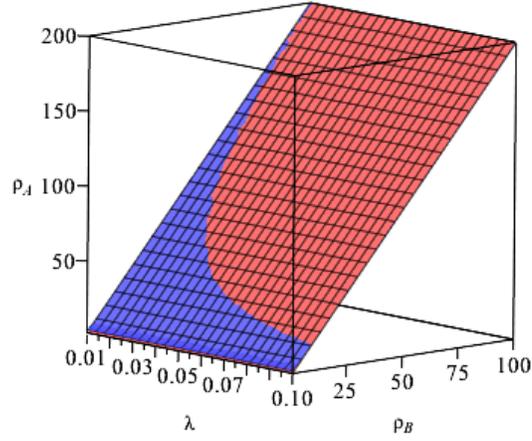}
\end{center}
\caption{We added the red curved surface $3\lambda^{2}+4\rho_{A}\lambda+4\rho_{B}\lambda-4\rho_{A}^{2}+8\rho_{A}\rho_{B}-8=0$
 to FIG.\ref{Fig.2}. Above this red curved surface, the gauge fixed point is stable. Note that part of the 
red curved surface is hidden under the blue surface
and the same applies to the blue curved surface.}
\label{Fig.3}
\end{figure}

\begin{figure}[H]
\begin{center}
\includegraphics[width=8cm]{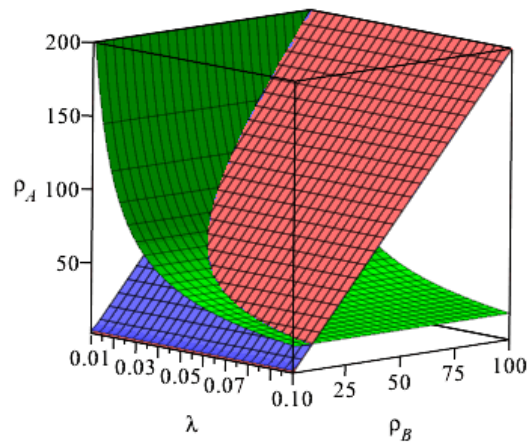}
\end{center}
\caption{The green curved surface $\lambda^{2}+2\rho_{A}\lambda-4 =0$ is added to the FIG.\ref{Fig.3}. 
The gauge fixed point exists above this surface.}
\label{Fig.4}
\end{figure}

\begin{figure}[H]
\begin{center}
\includegraphics[width=8cm]{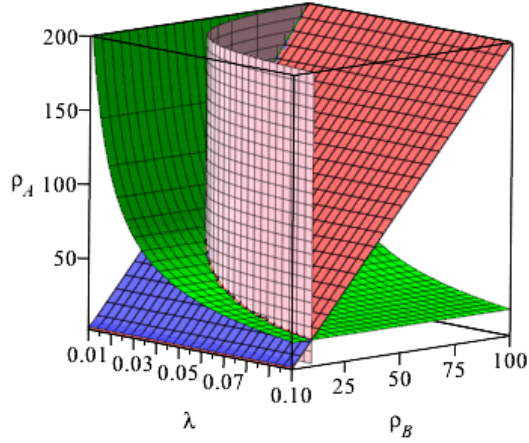}
\end{center}
\caption{The pink curved surface $\lambda^{2}+2\rho_{B}\lambda-2 =0 $ is added to the FIG.\ref{Fig.4}.
Below the red surface and above the blue surface, there are four fixed points of which only the hybrid fixed point is an attractor. 
Above the red surface and  right of the pink surface, the gauge fixed point  is an attractor and 
the isotropic and two-form fixed points are saddle points.
Left of the pink surface and above the green surface, the gauge fixed point is an attractor and the isotropic fixed point is a saddle point.
Below the blue surface and  above the green surface, the two-form fixed point is an attractor and the isotropic and the gauge fixed points
are saddle points. Below the green surface and right of the pink surface, the two-form fixed point is an attractor and the isotropic fixed point is a saddle point.
Below the green surface and  left of the pink surface, only the isotropic fixed point exists and is stable.}
\label{Fig.5}
\end{figure}

Now, we are in a position to discuss the dynamical structure of phase space.
At the end of the day, we will see that there is only one stable fixed point and the others are saddle points if any.

First, we classify solutions in the space of coupling constants.
In FIG.\ref{Fig.2},  we plotted the blue  curved surface satisfying the relation 
\begin{eqnarray}
\lambda^{2}+2\rho_{A}\lambda-\rho_{B}\lambda+2\rho_{A}\rho_{B}-4\rho_{B}^{2}-4 =0 \label{blue}
\end{eqnarray}
in the parameter space $(\lambda , \rho_A , \rho_B )$. 
 Below this surface, the two-form  fixed point is stable. Conversely, above this curved surface, the two-form  fixed point is unstable.
Next, in FIG.\ref{Fig.3},  we added the red curved surface 
\begin{eqnarray}
3\lambda^{2}+4\rho_{A}\lambda+4\rho_{B}\lambda-4\rho_{A}^{2}+8\rho_{A}\rho_{B}-8=0 \label{red}
\end{eqnarray}
 to FIG.\ref{Fig.2}.  Above this red curved surface, the gauge fixed point is stable.  On the other hand, below this red surface, the gauge fixed point is unstable.
Note that part of the red curved surface is hidden behind the blue surface, and vice versa.
In fact,  below the red surface  (\ref{eq62}) and above the blue surface   (\ref{eq61}) , there exists the 
hybrid fixed point which we believe stable.
In FIG.\ref{Fig.4}, the green curved surface $\lambda^{2}+2\rho_{A}\lambda-4 =0$ has been added to  FIG.\ref{Fig.3}. 
Note that the gauge fixed point exists above this surface (see Eq.(\ref{eq52}))  where isotropic fixed point is unstable.
In FIG.\ref{Fig.5}, the pink curved surface $\lambda^{2}+2\rho_{B}\lambda-2 =0 $ has been added to  FIG.\ref{Fig.4}.
In the right of the pink surface, there exists the two-form fixed point.
Here, we obtain the intersection curve of the green and the pink surface as
\begin{eqnarray}
  \rho_A &=& \frac{2}{\lambda} - \frac{\lambda}{2}   \label{green} \ ,\\
 \rho_B &=& \frac{1}{\lambda} - \frac{\lambda}{2}    \label{pink}  \ .
\end{eqnarray}
Remarkably, substituting these Eqs.(\ref{green}) and (\ref{pink}) into Eqs. (\ref{blue}) and (\ref{red}), we see these four surfaces
intersect exactly at the same curve defined by  Eqs.(\ref{green}) and (\ref{pink}).

\begin{figure}[!h]
\begin{center}
\includegraphics[width=9cm]{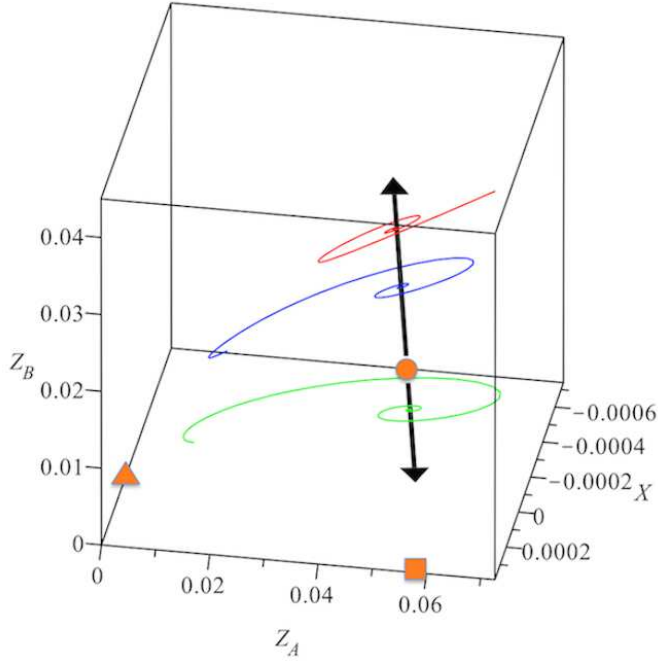}
\end{center}
\caption{The coupling constants are set to be $\lambda$=0.1, $\rho_{A}$=60.1, $\rho_{B}$=30.
 The number of e-foldings is run from 0 to 10.
The initial conditions are taken as ($X=0.00015, Y=0.01, Z_{A}=0.07, Z_{B}=0.045$) for the red trajectory, 
($X=0.0001, Y=0.01, Z_{A}=0.02, Z_{B}=0.02$)  for the blue trajectory, and
($X=0.0002, Y=0.01, Z_{A}=0.015, Z_{B}=0.01$)  for the green trajectory. 
The allow represents the direction of eigenvector corresponding to the slow mode.
The isotropic, gauge, and hybrid fixed points are represented by triangle, square, and circle, respectively. 
The two-form fixed point $(X,Z_A,Z_B)$=(-0.000740, 0, 0.0816) 
 is located outside of the range.}
\label{Fig.6}
\end{figure}

We can now summarize the results. 
Below the red surface and above the blue surface, there are four fixed points of which only the hybrid fixed point is an attractor
and other fixed points are saddle. 
We see that,  above the red surface and the right of the pink surface, the gauge fixed point  is an attractor and 
the isotropic and two-form fixed points are saddle points.
In the left of the pink surface and above the green surface, the gauge fixed point is an attractor and the isotropic fixed point is a saddle point.
Below the blue surface and the above the green surface, the two-form fixed point is an attractor and the isotropic and the gauge fixed point
are saddle points. Below the green surface and the right of pink surface, two-form fixed point is an attractor and isotropic fixed point is a saddle point.
Below the green surface and the left of pink surface, only isotropic fixed point exists and is stable.
Thus, we can conclude that only one fixed point is stable and others become saddle points if any.

\begin{figure}[!h]
\begin{center}
\includegraphics[width=11cm]{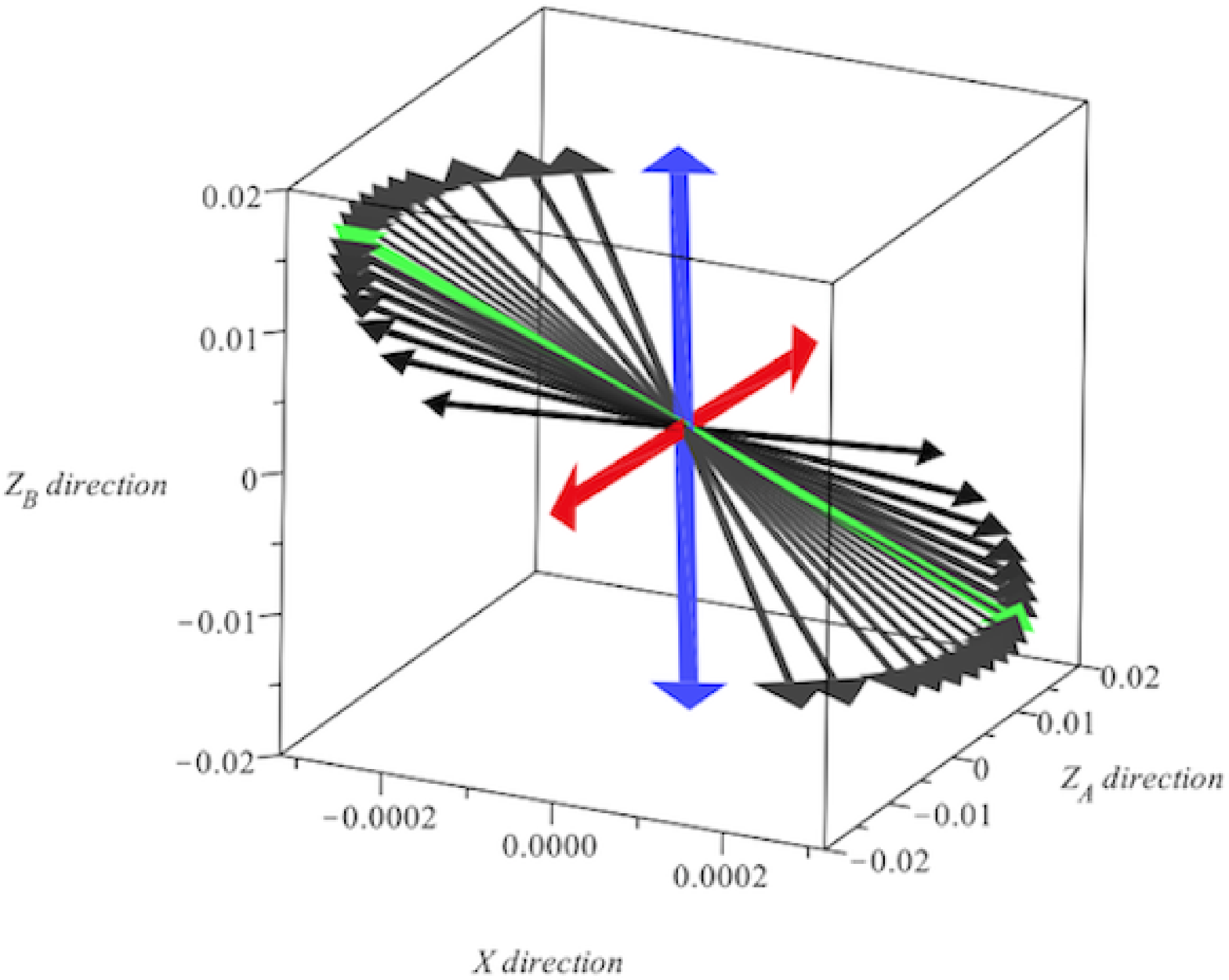}
\end{center}
\caption{The direction of slow mode from each hybrid fixed points. 
Here, $\rho_{B}=30,\lambda=0.1$ are fixed. Then, we change $\rho_{A}$ in the region where hybrid fixed point is stable.
When $\rho_{A}$ takes the lower limit, slow mode is parallel to the  $Z_{A}$ direction. This is represented by the red vector.
When we increase $\rho_{A}$, the eigenvector varies continuously. The vector at $\rho_{A}=2\rho_{B}$ is represented by the 
green vector. Finally, when $\rho_{A}$ reaches the upper limit, the vector becomes  parallel to $Z_{B}$ direction. 
This is represented by the blue vector.}
\label{Fig.vec}
\end{figure}

\begin{figure}[!h]
\begin{center}
\includegraphics[width=9cm]{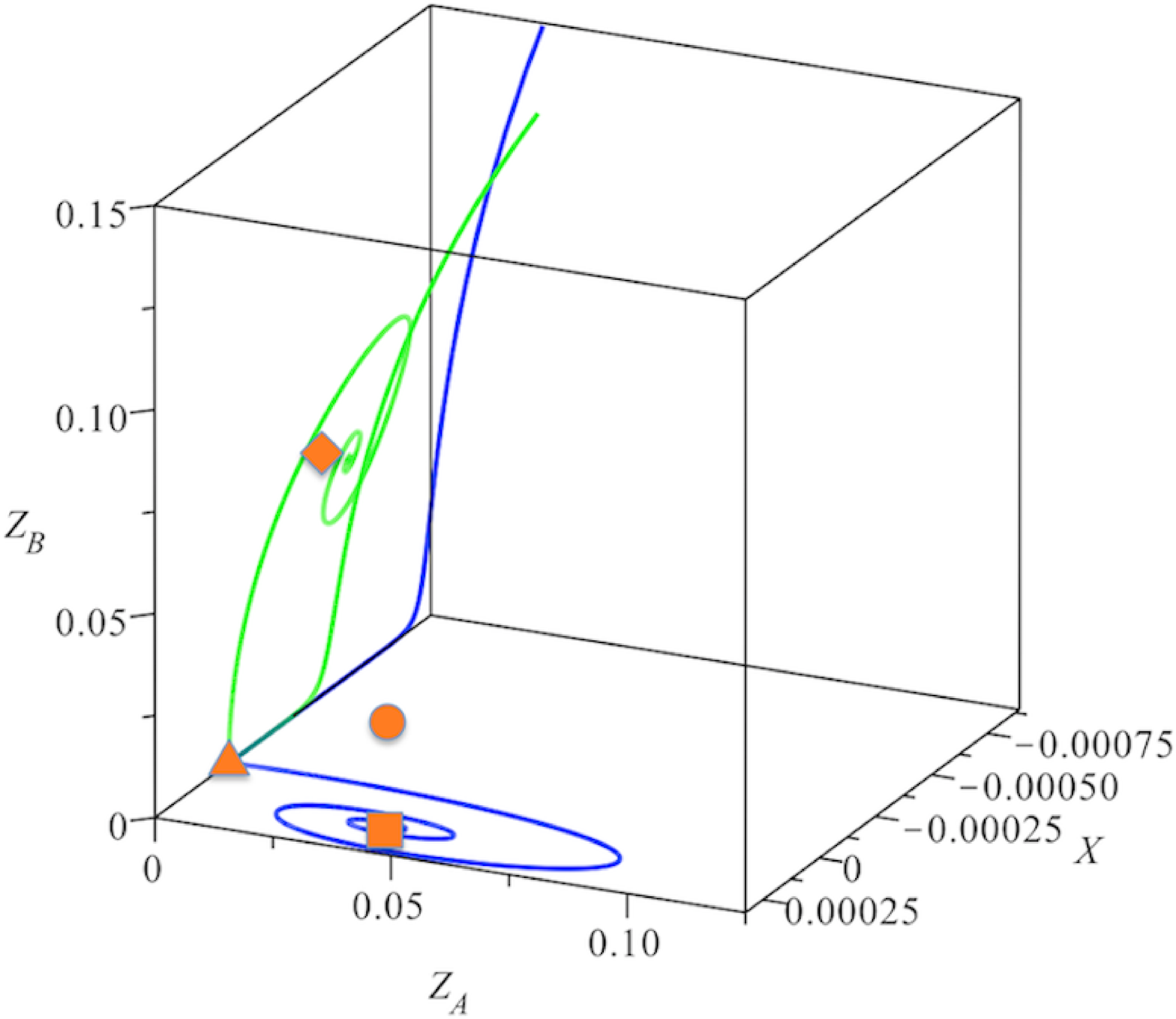}
\end{center}
\caption{We set coupling constants $\lambda$=0.1, $\rho_{A}$=140.1, $\rho_{B}$=70 and run the number of e-foldings from 0 to 15.
The initial values are chosen as ($X=0, Y=0.01, Z_{A}=0.9, Z_{B}=0.9$) for the blue trajectory, and 
($X=0, Y=0.01, Z_{A}=0.7, Z_{B}=0.7$) for the green trajectory.
The isotropic, gauge, two-form, and hybrid fixed points are represented by triangle, square, diamond, and circle, respectively.}
\label{Fig.7}
\end{figure}

So far, we have investigated the stability of fixed points. 
Now, we know  one is an attractor and the others (if any) are saddle.
The trajectory should be very complicated due to the saddle points.
Any trajectory should approach to the attractor point. However, depending on the initial condition, there are several transient
periods when the trajectory is staying for some time near the saddle point.
Here, we show some trajectories for various initial conditions. 
In FIG.\ref{Fig.6},  each trajectory seems to approach an attractor, however, the convergent points are slightly different
depending on the initial conditions. 
This comes from the fact that  the first eigenvalues in Eqs. (\ref{other-set0})  $\sim$ (\ref{other-set}) are small. 
Therefore, the trajectories are still on the way to the fixed point. 
Ultimately, they should converge to the fixed point, but it does not reach the fixed point even at the number of e-foldings 100.
Hence, in reality, we see various anisotropic inflation.
%red
We investigated the direction of the slow mode numerically (FIG.\ref{Fig.vec}).
In practice, the direction of the slow mode through each hybrid fixed point becomes an attractor.
In FIG.\ref{Fig.vec}, we fixed $\rho_{B}$ and changed $\rho_{A}$ in the region where the hybrid fixed point is stable.
We also fixed $\rho_{A}$ and changed $\rho_{B}$, and we got a similar graph, that is , 
the direction of the slow mode is determined by the relative ratio between $\rho_{A}$ and $2\rho_{B}$.
So FIG.\ref{Fig.vec} represents universal behavior of the slow mode. 
%red
Other interesting trajectories are plotted in FIG.\ref{Fig.7}. 
In this case, at first, $Z_{A}$ and $Z_{B}$ fall to zero, which means the universe becomes isotropic. 
Next, each trajectory leave the isotropic saddle point and goes to another saddle point where the universe is anisotropic. 
Ultimately, they will converge to the stable fixed  point where the universe undergoes a different anisotropic expansion.
As the examples in FIG.\ref{Fig.6} and FIG.\ref{Fig.7} tell us, we can see that the feature of anisotropic inflation is
 somewhat influenced by the initial conditions. This can be understood from the fact that the eigenvalues of unstable modes in  
(\ref{eq111}) and (\ref{eq110}) are small. Moreover, if we  choose a different set of coupling constants, 
we can design a more variety of anisotropic inflation. 

%red
Having said the initial condition dependence, one may wonder if the models can be predictable.
From the Fig.\ref{Fig.vec}, we see the direction of the slow mode depends on the relative ratio  between $\rho_{A}$ and $2\rho_{B}$.
Once we fixed the model parameters  $\rho_{A}$ and $\rho_{B}$, we can determine the eigenvector.
Moreover, an initial condition selects a point on this eigendirection.
Thus, we can read off the magnitude of $Z_{A}$ (energy density of gauge field) and  $Z_{B}$ 
(energy density of two-form field). Note that $Z_{A}$ and $Z_{B}$ are correlated because they are on the line.
 For making predictions, these values are sufficient.
Of course, since we cannot know the initial condition,  we can not predict each observable.
Nevertheless, we can falsify the model by checking the consistency conditions between observables
of the model. In this sense, our models have predictability.
%red

\section{Conclusion}

We studied inflationary universe from the point of view of cosmic democracy.
More precisely, we investigated phase space of anisotropic inflation in the presence of form fields, namely,
 a gauge field and a two-form field.
We sought exact power-law solutions when the potential of inflaton and the gauge kinetic functions are exponential type. 
We found new exact anisotropic inflationary solutions. 
Next, we cast the system into an autonomous system and found fixed points corresponding  to the power-law solutions.
We showed that only one fixed point could be stable and the others are saddle points if any. 
The coupling constants $\lambda,\rho_{A},\rho_{B}$ determined which one becomes stable. 
The result is summarized in FIG.\ref{Fig.5}.
In the case of hybrid fixed point, we found the convergence to the attractor is very slow.
As a consequence, it turns out that the dynamics of the anisotropic inflation depends on the initial conditions.
We also found trajectories which go around several saddle points before reaching the final attractor point.
We also demonstrated how the phase space of anisotropic inflation is fertile.

There are several applications to be considered.
Our results may have implication for generation of primordial magnetic fields~\cite{Kanno:2009ei}. 
It is worth investigating generation of magnetic fields in this context of cosmic democracy.
In the present model, an isotropic inflation can be realized in spite of the presence of
 the gauge field and the two-form field because both tend to produce opposite anisotropy
(This is similar to the multi-vector cases~\cite{Yamamoto:2012tq,Yamamoto:2012sq}.).
 It is intriguing to study observational signature of this special case.
In general, of course, we have anisotropic inflation. It is also interesting to look at observational predictions of hybrid models on the 
cosmic microwave background radiation. 
%red
{Depending on the solutions, the predictions would be different. For example, the shape of the anisotropy depends on the initial conditions.
However, that does not mean the model is not falsifiable. In fact, there are several consistency conditions between observables
which enable us to get information independent on initial conditions. 
%red
We leave the detailed study to future work.

From the perspective of designing anisotropic inflation, there are several directions to be investigated.
It is possible to extend anisotropic inflation to non-abelian models~\cite{Murata:2011wv},
 k-inflation models~\cite{Ohashi:2013pca}. We can also incorporate the parity violation in anisotropic inflation~\cite{bartolo}.
These are complementary to the effective field theory approach~\cite{Cannone:2015rra}.

\acknowledgements
AI would like to thank Kei Yamamoto for useful discussion.
This work was supported by  Grants-in-Aid for Scientific Research (C) No.25400251
 and Grants-in-Aid for Scientific Research on Innovative Areas No.26104708.

\end{document}